# Exception Handling in Multiparty Interactions

B. Randell
Department of Computing Science
University of Newcastle upon Tyne
NE1 7RU Newcastle upon Tyne - UK
brian.randell@ncl.ac.uk

A. F. Zorzo[*]
Faculdade de Informática
Pontifícia Universidade Católica do RS
90619-900 Porto Alegre - RS - Brazil
zorzo@inf.pucrs.br

**Abstract**

In designing distributed and parallel systems, there are several approaches for programming interactions in a multiprocess environment. Usually, these approaches take care only of synchronisation or communication in two-party interactions. This paper is concerned with more general concept: *multiparty interactions*. In a multiparty interaction, several executing threads somehow "come together" to produce an intermediate and temporary combined state, use this state as a well-defined starting point for some joint activity, and then leave this interaction and continue their separate execution. The concept of multiparty interactions has been investigated by several researchers, but to the best of our knowledge none have considered how faults in one or more participants of the multiparty interaction can best be dealt with. The goal of this paper is twofold: to show how an existing specification language can be extended in order to allow *dependable* multiparty interactions (DMIs) to be declared; and to present an object-oriented framework for implementing DMIs in distributed systems. To show how our scheme can be used to program a system in which multiparty interactions are more than simple synchronisations or communications, we use a case study based on an industrial production cell model developed by Forschungszentrum Informatik, Karlsruhe, Germany.

## 1 Introduction

Parallel programs are usually composed of diverse concurrent activities, and communication and synchronisation patterns between these activities are complex and not easily predictable. Thus, parallel programming is widely regarded as difficult: [Foster 1996], for example, says that parallel programming is "more difficult than sequential programming and perhaps more difficult than it needs to be". In addition to the normal programming concerns, the programmer has to deal with the added complexity brought about by multiple threads of controls: managing their creation and destruction and controlling their interactions via synchronisation and communication.

As in sequential programming, complexity in program development can be managed by providing appropriate programming language constructs. Language constructs can help both by supporting encapsulation so as to prevent unwanted interactions between program components and by providing higher-level abstractions that reduce programmer effort by allowing compilers to handle mundane, error-prone aspects of parallel program implementation [Foster 1996].

A mechanism that encloses multiple processes executing a set of activities together is called a *multiparty interaction* [Joung & Smolka 1996] [Forman & Nissen 1996] [Attie 1993] [Evangelist *et al.* 1989]. In a multiparty interaction, several executing processes somehow "come together" to produce an intermediate and temporary combined state, use this state to execute some joint activity, and then leave the interaction and continue their normal execution.

---

[*]Present address: Dept. of Computing Science, University of Newcastle upon Tyne.

There has been a lot of work in the past years on multiparty interaction, but most of it has been concerned with synchronisation, or handshaking, between parties rather than the enclosure of several programmed activities executed in parallel by the interaction participants. For example, specification languages like CSP, LOTOS, or programming languages like Ada, only deal with synchronisation between processes. However, the programmer designing a set of processes that are taking part in a cooperating activity is left with full responsibility for ensuring that it is just these processes that are involved in the activity, and that they do not interfere with, or suffer interference from, other processes that are not supposed to be involved.

In designing a language for multiparty interaction, one must make a trade-off between the implementation efficiency of the language versus its expressive power. There are several choices that can be made when designing a multiparty interaction construct. For example, [Evangelist *et al.* 1989] describes a set of properties for a multiparty interaction that serves as an interprocess communication primitive. These properties are:

- Pre-synchronisation: in synchronous multiparty interaction constructs the participants of an interaction must synchronise before the interaction commences, i.e., if one participant arrives, it has to wait until all participants of the interaction have arrived. The main effect of this property is to provide a consistent combined state before the interaction starts.

- Split bodies: each participant in the interaction has its own set of commands that is executed in parallel as part of the interaction. Multiparty interactions that do not have split bodies usually have just one block of code that is executed by only one participant of the interaction - a special case that will not be considered further here.

- Frozen initial state: the participants of the interaction view the combined state as frozen in the beginning of the interaction, until the end of the interaction, when all changes that were made take effect. From the point of view of participants that are not involved in the interaction, this property can be seen as an atomic change of state in the system. Such guarantee will avoid wrong information from being accessed by processes outside of the interaction.

Other features are included in certain language constructs [Forman & Nissen 1996] [Back & Kurki-Suonio 1988], and are related to the way the multiparty interaction is activated or terminated. These features are:

- Pre-conditions: some interaction mechanisms provide a guard to check the pre-conditions to execute the interaction, hence the need for having synchronisation upon entry. If the pre-condition is true, then the interaction can commence, otherwise the interaction is not executed.

- Post-conditions: an assertion after the interaction has finished to guarantee that a set of post-conditions has been satisfied by the execution of the interaction.

Additionally, [Joung & Smolka 1996] presents a list of choices that can be made in the design of language constructs for multiparty interaction. These choices include the following:

- Biparty vs. multiparty interactions: as the name indicates, biparty interactions involve only two participants, and multiparty interactions are not so limited, and instead typically involve several participants.

- Fixed vs. variable interactions: in the former, the set of participants of an interaction is fixed, i.e. they do not change every time the interaction is executed. In the latter the participants are variable, i.e. participants can be different each time the interaction is executed. (Fixed interactions are referred to as "zeroth-order" and variable ones as "first-order" interactions in [Joung & Smolka 1994].)

- Conjunctive vs. disjunctive parallelism: conjunctive parallelism allows a set of interactions to be executed simultaneously as an atomic unit, while disjunctive parallelism chooses, non-deterministically, one interaction to be executed from a set of possible interactions.

- Synchronous vs asynchronous execution in the underlying system: in synchronous systems every interaction has to execute one step of its computation at a time, while in asynchronous systems there is no such restriction.

Based on the above set of choices, [Joung & Smolka 1996] presents a detailed taxonomy of languages that have a multiparty interaction mechanism as a basic construct. What is of particular interest here is that they point out that the expressive power of a language that has a multiparty interaction as a basic mechanism is dependent on the way participants can enroll in an interaction. They identify four basic interaction constructs based on their support of multipartiness and of variable interactions: *channels, ports, gates,* and *teams*.

A **Channel** is a primitive for biparty communication. It is used as a communication link between two processes. The communication actually occurs only when both processes are ready to communicate. One example of channel usage is the input/output command in CSP [Hoare 1985]. The input command $P_i?y$ of processes $P_j$, which inputs a value from process $P_i$ into variable $y$, is complementary to the output command $P_j!x$ of process $P_i$, which outputs the value of expression $x$ to $P_j$. The joint execution of these commands is equivalent to the assignment of $x$ to $y$ ($y \leftarrow x$).

A **Port** is also a primitive for interaction between two processes. It is a mechanism for achieving variable interactions that define an activity involving two "roles". In a port-based interaction, a process does not know in advance with which process it is interacting. For example, in the readers-writers problem, a port can be defined with two roles, one for the readers and the other for the writers. Any reader process that needs a new value from a writer can *enrole*[1] into the reader's role, while any writer ready to output the role can enrole into the writer's role. One language that uses this kind of primitive is Ada, whose *rendezvous* represents a port-based interaction involving two roles, one assumed by a fixed callee and the other by callers.

A **Gate** is a multi-channel primitive that defines an interaction among a fixed number of processes. An example of a language that uses a gate as the way processes enroll in an interaction is LOTOS [Bolognesi & Brinksma 1987].

A **Team** has the same properties as a port, with a fixed number of roles, although the number of roles is not limited to two. A set of processes can jointly establish an instance of the team by filling all the roles. Examples of team primitives can be found in DisCo [Järvinen & Kurki-Suonio 1991] or in Multiway Rendezvous [Charlesworth 1987].

---

[1] In [Forman & Nissen 1996] *enrole* is used because they think a new word is necessary to express the enrollment of processes into roles.

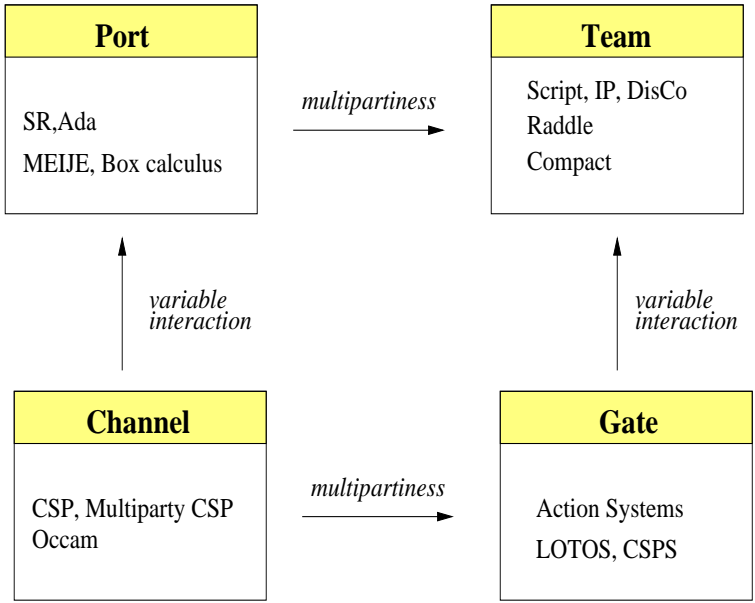

Figure 1: A Taxonomy of Languages for Multiparty Interactions

Figure 1 shows a possible taxonomy of a number of languages for multiparty interactions[2]. The figure represents the expressive power of the languages, where the team-based languages have the most expressive power and the channel-based languages have the least expressive power. This can be concluded from the following observations. Port-based languages can describe channel-based systems by assigning a port $p_{ij}$ to each pair of processes $P_i$ and $P_j$ such that only $P_i$ and $P_j$ have access to $p_{ij}$ and that they always execute the same role in that port. Similarly, since ports are the biparty equivalent of teams, team-based languages can also describe channel-based systems. Similar reasoning regarding channels and ports shows that team-based languages can describe gate-based systems. Given their greater expressive power, we have therefore chosen team-based interaction languages as the basis of the work we describe in this paper.

It should be noted that the team concept has largely been investigated as a specification language construct, whereas our interest in it concerns its utility not just within specifications but also for structuring and designing actual programs. Moreover, to date programming languages usually provide only two-party synchronisation mechanisms, such as the rendezvous in Ada. Although there has been a lot of research on multiparty interaction mechanisms, to the best of our knowledge none has considered the provision of features that would facilitate the design of multiparty interactions that are expected to cope with faults - whether in the environment that the computer system has to deal with, in the operation of the underlying computer hardware or software, or in the design of the processes that are involved in the interaction.

Because faults are expected to occur rarely during the execution of a program, programmers usually refer to them as *exceptions* [Cristian 1982]. To structure the handling of the exceptions, which may occur during the execution of a program, an exception mechanism is often provided in a programming language. Such a mechanism allows a programmer to make it explicit that the normal flow of a program has to be replaced by an exceptional flow whenever an exception is detected in that program.

---

[2]For further reading about the languages or algebraic models presented in Figure 1 refer to: CSP [Hoare 1985], Multiparty CSP [Joung & Smolka 1990], Occam [Hoare 1984], SR [Andrews *et al.* 1988], Ada, MEIJE [Simone 1985], Box calculus [Best *et al.* 1992], Action Systems [Back & Kurki-Suonio 1988], LOTOS [Bolognesi & Brinksma 1987], CSPS [Raman & Day 1984], Script [Francez *et al.* 1986], IP [Forman & Nissen 1996], Compact [Charlesworth 1987], Raddle [Forman 1986], and DisCo [Jårvinen & Kurki-Suonio 1991].

In this paper we use the term *dependable multiparty interaction* (DMI) for a multiparty interaction that provides facilities for:

- Handling Concurrent Exceptions: when an exception occurs in one of the bodies of a participant, if it is not dealt with by that participant, the exception must be propagated to all participants [Campbell & Randell 1986] [Romanovsky *et al.* 1996]. A DMI must provide a way of dealing with exceptions that can be raised by one or more participants. If several different exceptions are raised concurrently, then the DMI mechanism has to decide which exception will be raised in all participants.

- Assuring Consistency Upon Exit: a participant can only leave the interaction when all of them have finished their roles and the external objects are in a consistent state. This property guarantees that if something goes wrong in the activity executed by one of the participants, then all participants have an opportunity to recover from possible errors.

In view of our interest in dependability, and in particular fault tolerance, we adopt the use of pre- and post-conditions, which are checked at run-time. Regarding the remaining alternatives listed earlier regarding multiparty interactions, we have made the following design choices for our DMIs:

- Although the particular processes involved should be able to vary, their number in a given DMI should be fixed.

- The processes should synchronise their entry to and exit from the DMI.

- The DMI mechanism should ensure that as viewed from outside the DMI, its system state should change atomically, though inside the DMI intermediate states will be visible (c.f. the related "frozen state" property discussed above [Evangelist *et al.* 1989]).

- The way the underlying system executes a DMI can be synchronous or asynchronous.

The work on DMIs described in this paper makes use of a language that we have chosen because of its very high expressive power, and because it has several of the properties listed above as ones that we consider important for DMIs: DisCo [Jårvinen & Kurki-Suonio 1991]. DisCo is a specification language for reactive systems developed at Tampere University of Technology, Finland. It can be characterized as both action-oriented and object-oriented.

In this paper we show how *dependable multiparty interactions* can be described in a DisCo-like specification language, and how to implement DMIs in an object-oriented framework that provides support for DMIs in distributed applications. Section 2 presents the DisCo specification language, and uses it to describe a control system for a simplified version of an industrial production cell case study [Lötzbeyer & Muhlfeld 1996]. Section 3 presents a possible semantics for handling exceptions that may be raised during an action and discusses how to extend DisCo in order to specify such exceptions. This uses, as a running example, various aspects of a much more complete and realistic version of the industrial production cell case study, much of whose complexity comes from the need to deal with a large number of possible error situations. Finally, Section 4 presents an object-oriented framework for implementing DMIs.

## 2 Distributed Cooperation - DisCo

DisCo is a specification language based on the Action Systems approach, in which a designer has to concentrate on the interactions between components rather than on the components themselves [Back & Kurki-Suonio 1988]. An action system consists of a set of state variables and a set of actions. Each action is composed of a guard and a body. A guard is a boolean expression involving state variables, and the body is a set of commands to change the state of the variables. DisCo extends Action Systems into the object-oriented paradigm.

A program specification in DisCo is composed of a set of two basic components: *objects* and *actions*. Objects are instances of classes and are the means of representing the global state of a system. Objects are called participants in a DisCo action. Actions are the only units of execution in DisCo. They enclose a sequence of state transformations, and are the only means by which the state of an object can change. Actions are executed nondeterministically, and the execution of an action is atomic, meaning that once the execution of an action has started, it cannot be interrupted or interfered by other actions.

In this section we show how to apply DisCo to specify the basic elements of a production cell case study that was developed at the Forschungszentrum Informatik (FZI), Karlsruhe, Germany. This "Production Cell II" case study involves the control of a set of machines that are used in combination in order to achieve a particular mechanical production process. We defer to later sections of this paper consideration of the problems that were a major motivation behind the creation of this particular case study, namely how to deal with the many and varied ways in which the various machines and their associated sensors and actuators might fail, and first concentrate simply on the problems of how to ensure that the machines cooperate properly, i.e., do not interfere with each each other.

**Case Study - FZI Production Cell II**

The FZI Production Cell II case study [Lötzbeyer & Muhlfeld 1996] that we used for the work discussed in this paper is an extension of a production cell case study described in [Lewerentz & Lindner 1995], which is a model based on an actual industrial installation in a metal-processing plant in Karlsruhe, Germany. The Production Cell II case study describes a production cell composed of six devices: two conveyor belts – a feed belt and a deposit belt, an elevating rotary table, two presses, and a rotary robot that has two orthogonal arms. The state of devices is reflected by sensors that provide information about their position. Each device has a set of actuators that are used by a control program to change their state.

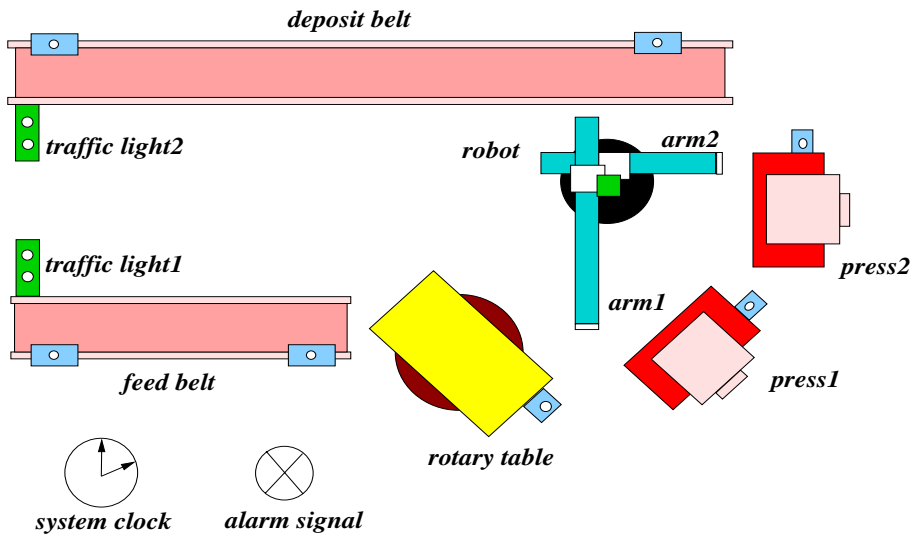

Figure 2: The FZI Production Cell II

A complete production cycle of a metal plates is as follows: *i*) if the traffic light at the beginning of the feed belt is green, then a metal plate can be added on the feed belt; *ii*) the feed belt conveys the metal plate to the elevating rotary table; *iii*) the table rotates and elevates to the position where the robot can grab the plate; *iv*) the first arm of the robot grabs the plate and places it into a free press (press1 or press2); *v*) the chosen press forges the plate; *vi*) the second arm of the robot removes the forged plate from the press and places it on the deposit belt; *vii*) if the traffic light at the end of the deposit belt is green, then the plate is carried out of the production cell by the deposit belt. Figure 2 shows all devices cited in this paragraph. (The full system includes numerous actuators and further sensors.)

### DisCo Objects

The state of a system in DisCo consists of a set of objects. Objects are composed of attributes that can represent simple values, such as integers or boolean values, or sets of simple values. Object attributes can also include finite state machines and references to other objects. Each object can contain as many groups of attributes as needed. An object can be composed also of other objects, e.g. the Robot object has two Arm objects (see code below). A star (*) in front of one element of a finite state machine means that that is the initial value from that state group.

Here are some examples of DisCo class definitions for the FZI Production Cell II case study:

```
class Arm is
    state *pos_ret, pos_ext, pos_mv, pos_out;
    state *free, loaded;
end;

class Robot is
    arm1, arm2: Arm;
    state *table, load_press1, load_press2, unload_press1, unload_press2, deposit_belt;
end;

class Table is
    state *pos_feedbelt, pos_belt2robot, pos_robot, pos_out;
    state *lower, upper;
    state *free, loaded;
end;

class TrafficLight is
    state *green, red;
end;
```

In the above examples, when an object from class Table is created, its initial state is: pos_feedbelt, lower, and free, meaning that the table is pointing towards the feed belt and is in its lower position, and there is no plate on the table. In summary, the table is ready to be loaded.

### DisCo Actions

Actions are the execution entities in DisCo. (Such actions are, as shown in Figure 1, teams; more specifically they are multiparty interactions with a fixed number of variable participants.) Each action has a guard, which is a predicate, and a body. When the guard is true for a collection of potential participants, the action is said to be enabled. Objects are the participants in a DisCo action and assume a role in the action when the action is enabled. The body of an action consists of one sequential set of assignments and conditional statements which can refer only to the participants and parameters of the action.

In DisCo, there is no concept of process threads; actions are executed when their guard (**when** condition) is true and the objects involved in the action are not being used in another action. If two actions that use a same object can be activated at the same time, then only one of them will be executed. (The choice is non-deterministic.)

Before we give an example of actions in DisCo, let us consider how actions could be used for controlling the interactions between devices in the Production Cell II case study. Figure 3 portrays the way in which we have chosen to use actions so as to structure the controlling of a sequence of operations between devices. Each action encloses a set of devices that must interact in a coordinated fashion to satisfy the requirements that the case study defines. If two such actions are shown as overlapping, this indicates that they must not be performed in parallel because they both involve the same device. The semantics of DisCo will guarantee this. For example, the *LoadTable* action cannot be executed in parallel with the *UnloadTable* action because both actions involve the table object, and the table can participate in only one of them. Our controlling software for FZI Production Cell II is thus composed of 12 actions as shown in the figure.

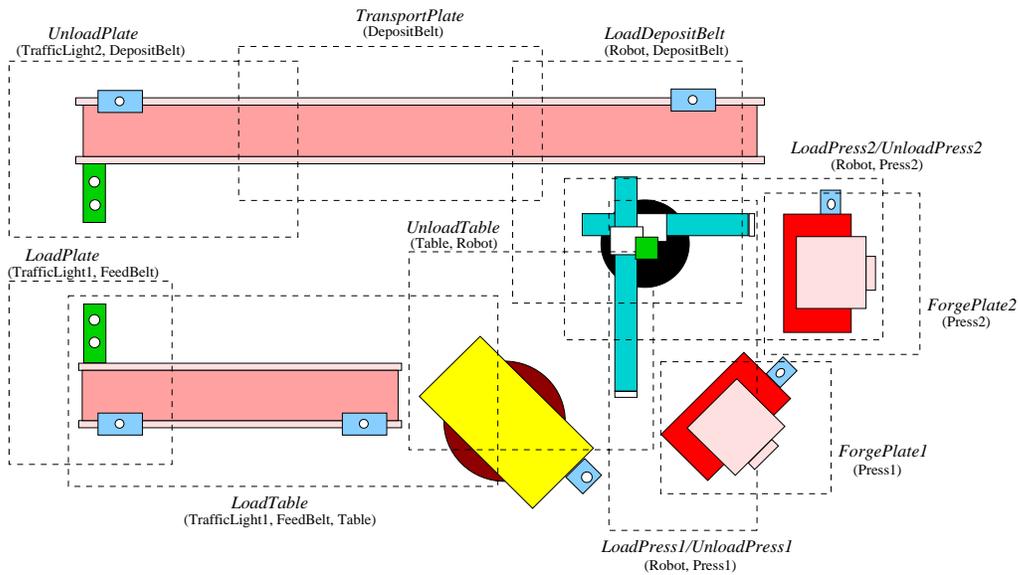

Figure 3: DMIs in FZI Production Cell II

The description below shows how the *LoadTable* and *UnloadTable* would be described in DisCo. It shows how the table in Production Cell II would cyclicly activate the loading and unloading of the table with metal plates. The loading of the table is an example of a multiparty interaction that is executed by three parties: the traffic light at the beginning of the feed belt; the feed belt; and the table. The change of state value in an object is represented by → new-state-value, e.g. → tf1.green means that the traffic light changes its state to the green state. The unloading of the table involves just the table and the robot.

**action** LoadTable **by** tf1:TrafficLight; fb:FeedBelt; t:Table **is**
**when** tf1.red ∧ fb.loaded ∧ t.free **do**
    → t.lower;
    → t.pos_feedbelt;
    → fb.free;
    → t.loaded;
    → tf1.green;
**assert** t.lower ∧ t.pos_feedbelt ∧ t.loaded ∧ fb.free ∧ tf1.green
**end**;

```
action UnloadTable by t:Table; r:Robot is
when t.loaded ∧ r.arm1.free do
    → t.free;
    → r.arm1.loaded;
    → r.table;
assert t.free ∧ r.arm1.loaded ∧ r.table
end;
```

In DisCo, an action's pre-condition is specified via a guarded command, which is executed before the action commences (**when** clause). DisCo also allows a designer to specify assertions (**assert** clauses) in the code of an action. Assertions are checked at the place they occur. If an assertion is not true, then the system is stopped. Action post-conditions can thus be specified by placing an assertion at the end of the action code (see code above).

## 3   Adding Exception Handling to DisCo

The full specification of Production Cell II calls for a controlling system that satisfies the following requirements:

- *Safety*. Device collisions and the dropping of plates must be avoided. Plates must keep a safe distance from each other. Device mobility must be restricted appropriately.
- *Liveness*. Any metal plate added into the cell via the feed belt must eventually leave the cell via the deposit belt and have been forged.
- *Fault Tolerance*. When a failure occurs, it should be detected and the system should be stopped in a safe state if possible. After recovery from the failure, the system should resume operation from this safe state.
- Other requirements, such as *flexibility* and *efficiency*, may be taken into account as long as they do not conflict with the above ones.

In Section 2 we presented a set of actions that would enclose the interactions between devices every time a metal plate was being passed from one device to another. The way actions were specified in Section 2 suffices for guaranteeing the safety and liveness requirements of the case study. (A similar approach was presented for the Production Cell I case study in [Zorzo *et al.* 1999] and its formal verification described in [Canver 1997].)

In this section we describe a possible semantics for handling exceptions in DisCo, and extend the DisCo language to be able to specify exceptional behaviour based on these semantics, so as to be able to specify the FZI Production Cell II fault tolerance requirements. We in fact use the exception handling semantics proposed for the Coordinated Atomic (CA) action [Xu *et al.* 1995] [Randell *et al.* 1997] mechanism. (A CA action is a multiparty interaction mechanism for coordinating and ensuring consistent access to objects in the presence of concurrency and potential faults.)

**Exception Handling Semantics**

The CA action mechanism we use involves multiple roles that must agree about the CA action outcome. There are four possible kinds of outcome: normal, exceptional, abort and failure. If a CA action does not terminate normally (i.e. without an exception), then each role must *signal* an exception to the enclosing action indicating the outcome. The roles should agree about the outcome so each role should *signal* the same exception. Note that there is a distinction between an exception being *raised* within an action and an exception being *signalled* by an action to its enclosing action. If an action terminates by *signalling* an exception, then that exception is *raised* within the enclosing action and this triggers the process of exception handling within that action.

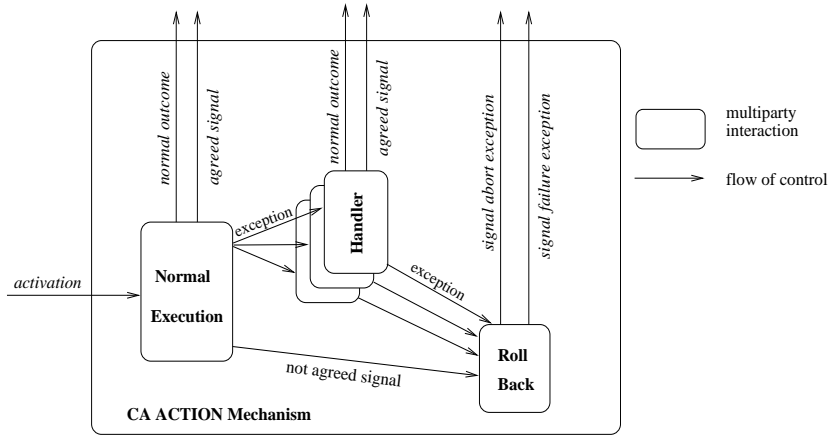

Figure 4: CA Action Mechanism

Figure 4 shows how the CA action mechanism is itself built using basic multiparty interactions. In the figure, rounded boxes represent multiparty interactions, arrows represent the flow of control, and the enclosing rounded box represents the CA action mechanism. As we can see in the figure, CA actions are implemented with three basic types of multiparty interactions: one interaction for the normal execution of the CA action; a set of multiparty interaction handlers for dealing with the exceptions that may be raised during the normal execution of the CA action; and a third interaction that deals with the rolling back process of the CA action. The key idea for handling concurrent exceptions is to build a DMI mechanism, such as the CA action, out of basic multiparty interactions by chaining them together appropriately, so that each basic multiparty interaction in the chain is the exception handler for the previous basic multiparty interaction in the chain.

The order in which the basic multiparty interactions shown in Figure 4 are executed is driven by the following CA action mechanism rules:

1. Each role in a CA action may terminate normally or exceptionally by signalling an exception to the enclosing CA action. The roles should agree about the outcome of the action. However, a role may also signal an abort exception or a failure exception to indicate that the action should abort or fail (see 7). In the figure this rule is represented by the *normal outcome* and *agreed signal* lines.

2. If the roles do not agree about the outcome, then the underlying CA action support mechanism attempts to abort the action by undoing its effects on external objects. This rule is represented in the figure by the *not agreed signal* line from the normal execution interaction to the roll back interaction.

3. If the abort is successful, then an abort exception is signalled to the enclosing CA action; otherwise, a failure exception is signalled. (See *signal abort exception* and *signal failure exception* lines in the figure.)

4. If an exception is raised during the normal execution of a CA action, then control is passed to the corresponding exception handler for each role. If two or more exceptions are raised concurrently, then a process of exception resolution must take place first.

5. If an exception is raised during the execution of an exception handler, then the underlying CA action support mechanism will attempt to abort the action (see 3) or else signal a failure exception to the enclosing action.

6. Once exception handling begins within a CA action, it is not possible to resume normal execution of the CA action but it is possible for the exception handlers to terminate normally or exceptionally, depending on the extent to which error recovery is successful. However, all roles must still agree about the outcome (see 2).

7. A role may signal abort or failure at any time to indicate that error recovery is not possible and the action must abort or fail. For the purposes of determining the outcome, failure takes precedence over abort which takes precedence over every other exception that can be raised internally.

8. For a given action, exception handlers are provided only for exceptions that are raised internally within that action. Exceptions that are signalled by an action are handled at the level of the enclosing action. Thus an action cannot provide an exception handler for its own abort or failure exceptions.

9. If an action terminates by signalling an exception to its enclosing action, then this triggers the process of exception handling in that action (see 5).

**Extending DisCo**

The first step in extending DisCo is to include means for the specification of exceptions that can be raised while changing the state of an object. This is introduced in the declaration of a class in DisCo. In the code below we extend the definition of objects by adding the clause **exceptions**. In this clause, the list of different types of exception that can be raised by that object are declared. We assume that only one exception can be raised per object at runtime. This is also the case where an object is composed of other objects.

```
class Table is
    state *pos_feedbelt, pos_belt2robot, pos_robot, pos_out, pos_err;
    state *lower, upper;
    exceptions lower_sensor, upper_sensor, plate_sensor, angle;
end;
```

Note that we have now included, in our example, the state pos_err among the possible states of Table - this is intended to represent the situation in which a possible fault in the table during the loading or unloading process, i.e., while the state of the Table object is changing from free to loaded, or vice-versa. Similar such states are also needed for the Arm and Robot classes.

The second extension we have made to DisCo is the definition of concurrency in the specification of an action. This helps us in representing (specifying) a real world in which several activities can happen in parallel during an interaction. In the example below, we have included several **by** clauses, one for each role in the action. All **by** clauses are executed in parallel as soon as all participant objects are available and the pre-condition (**when** clause) is true.

Specifications of exceptions are given after the specification of an object's possible state changes. Every time an object changes its state, an exception associated with that state transformation may occur. This is declared by including two colons (::) followed by an exception (or list of exceptions).

```
action LoadTable is
when tf1.red ∧ fb.loaded ∧ t.free do
by tf1:TrafficLight;
    → tf1.green;
by fb:FeedBelt;
    → fb.free :: fb.stuck;
by t:Table
    → t.lower :: t.lower_sensor;
    → t.pos_feedbelt :: t.angle;
    → t.loaded :: t.plate_sensor;
assert t.lower ∧ t.pos_feedbelt ∧ t.loaded ∧ fb.free ∧ tf1.green :: post-condition
end;
```

In the above example, when attempting to change the state of the table to pos_feedbelt, the table rotary motor can fail causing the table to be stuck. Such an occurrence is represented by raising the t.angle exception. As mentioned in Section 2, in DisCo if an assertion is not true, then the system stops. We have extended DisCo in order to allow exceptions to be specified which will be raised if an assertion is not true at the moment it is checked. In the above code, if the assertion, which represents the post-condition for LoadTable, is false, then the post-condition exception is raised.

Associated with an action, we have introduced the notion of a **handling action** that is executed when an exception is raised in an action of the same name. This handling action is responsible for bringing the objects of that action to a state that reflects the actual situation of the system. For example, if the angle of the table has failed, then the only valid state for the angle is pos_error. (This could not be represented in DisCo without the extensions we have applied.)

```
handling action LoadTable for t.angle is
by tf1:TrafficLight;
    → tf1.green;
by fb:FeedBelt;
    → fb.loaded;
by t:Table
    → t.free;
    → t.pos_error;
end;
```

Each action can also have associated with it a handling action for any exception that does not have its own specific handler. This handling action is defined without the **for** clause. One use for this special handling action is, for example, to implement a roll back action as shown in Figure 4.

In this section we have simply indicated how DisCo was extended in order to give a designer the tools for specifying how exceptions are handled in the system. We have not discussed the extension of the execution model of DisCo in order to allow action nesting. We have also not attempted to provide a formal treatment of our extension of DisCo, though this has in effect been done in a paper [Canver *et al.* 1998] which describes the use of model-checking to validate our Production Cell II program. A complete description of a language for DMIs is being prepared and will be published shortly in [Zorzo 1999].

# 4 DMIs in OO Languages

Specification languages, such as DisCo, are concerned with *what* is desired rather than *how* is is to be implemented [Jårvinen & Kurki-Suonio 1991]. In this section we present a generic framework for actually implementing DMIs, such as might be specified in the above extended version of DisCo, in object-oriented programming languages.

The framework described here is implemented in Java. A Java RMI ORB is used to distribute the objects of a DMI. The complete description of the classes for this framework can be found in [Zorzo & Stroud 1999]. This framework has been used for implementing the controlling software for Production Cell II presented in Section 2.

The framework we use is composed of 4 Java classes:

- `Role class`: each instance object of this class hosts a set of operations for one of the participants of the DMI.
- `Manager class`: a set of manager objects is responsible for keeping track of the components of the DMI, managing synchronisation of participants, testing the pre- and post-condition for the DMI, and deciding upon which exception is to be handled by all the participants of the DMI.
- `SharedObject class`: shared objects used for cooperation between the participants.
- `ExternalObject class`: external objects carry the state of the system in and out of the interaction. These objects have to provide some kind of transactional semantics.

Each instance object from the above classes can be distributed throughout a computer network. A DMI is programmed by grouping these objects in several sets. Each DMI is composed of: one set of objects from the above classes for interactions when there is no failures, i.e. *basic interactions*, and several sets of the above objects for dealing with each of the possible exceptions that may be raised during the execution of the interaction (either during the basic interaction or during an *exception handling interaction*).

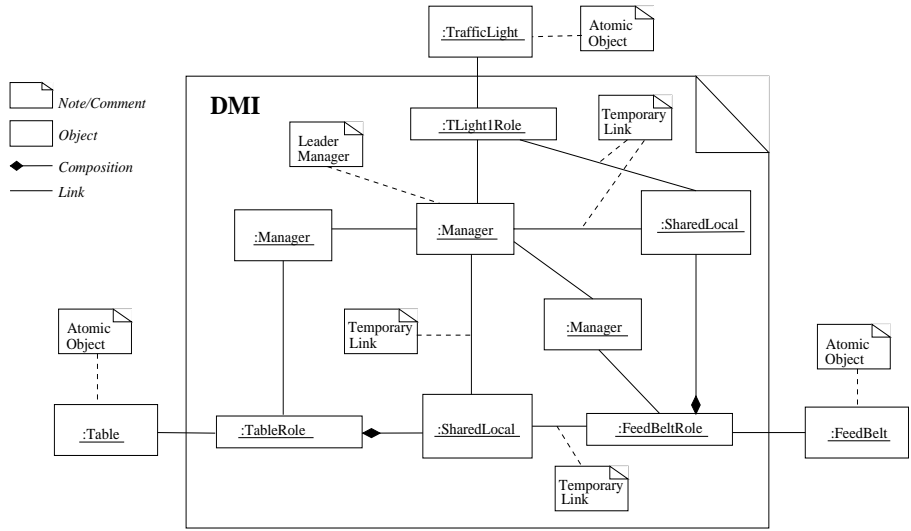

Figure 5: Links between the objects of the LoadTable DMI (in UML)

Figure 5 shows, using the UML notation [Fowler 1997], how the objects of DMI are organized in the Application Programming Interface (API) for the LoadTable action in the FZI Production Cell II described in Section 2. (Each of the objects in the figure could be distributed in a different host). Notice that all managers are connected to a special manager called the leader, and that the only manager that has a link to shared objects is the leader manager. External objects can be accessed (competitively) by other multiparty interactions. This access is serialized by the external objects in order to guarantee that only one multiparty interaction is accessing each object at a time, hence the atomic object notes in the figure. Some links in the figure exist only during the execution of the DMI. When the DMI commences, these temporary links are established. Those links are broken when the DMI terminates.

To program a new DMI using this framework, the first step is to define a new class that extends the `Role` class for each party in the interaction. The extended `Role` class has to redefine at least one method: the `body` method. This method will contain the set of operations that will be executed by the participant that activates the role. Upon creation each `Role` has to be informed about the manager that will be managing this role. A manager that 'controls' a `Role` object is an instance of the `Manager` class. (The `Manager` class provides a basis for coordinating the participants in a multiparty interaction.)

The managers of all roles will compose the controlling body of the interaction. Each manager upon creation is informed of which manager will act as the *leader* in the interaction. The leader is the responsible for controlling protocols for synchronisation between managers, for the exception resolution algorithm, and for keeping information about the shared objects. Every manager is a potential leader in our framework, avoiding a possible single failure point, if the host of the leader crashes.

## 4.1 Exception Handling

By default, the API we use to implement DMIs provides a built-in exception handling mechanism based on [Romanovsky *et al.* 1996]. This mechanism works as follows. When a role raises an exception, its manager is notified of that exception. The role's manager then informs the leader which interrupts all roles that have not raised an exception. After all roles have been interrupted or have notified the leader manager of an exception, an exception resolution algorithm is executed by the leader. This algorithm tries to identify a common ancestor[3] exception to all the raised exceptions. When such exception is found, the leader informs all the managers about that exception and an exception handling interaction (with the same features as a normal interaction) is activated. If there is no interaction handler for that exception, a handler for the highest level exception (`Exception` class) is tried. If there is no handler even for `Exception`, then the exception is passed (signalled) to the enclosing interaction.

Figure 6 shows a possible scenario in which two exceptions are raised during `LoadTable` DMI. Two roles, `FeedBeltRole` and `TableRole` raise exceptions `FeedBeltStuckException` and `TableAngleException` respectively (step **1** in the figure). These exceptions are caught by the role managers that inform the leader about these exceptions (step **2**). The leader then detects that `TLight1Role` is still executing and interrupts the thread executing that role (step **3**). An `InterruptedException` is therefore raised from the manager of `TLight1Role` informing the leader that the role has been interrupted successfully (in this case the manager of `TLight1Role` and the leader are the same) (step **4**). The leader then decides which exception has to be handled: exception `FeedBeltTableException` in our example. Exception `FeedBeltTableException` is sent to all managers of the DMI (step **5**) that will activate the roles in an exception handling interaction that will deal with exception `FeedBeltTableException` (step **6**). The set of managers and roles in the handler will then execute in the same way as if they belonged to a normal interaction.

In the event of one of the managers or one of the roles crashing, the managers communicate with each other and decide to raise a `CrashedManagerException` or a `CrashedRoleException` exception. If the manager that has crashed was the leader, then a new leader will be chosen by the managers that are still running. `CrashedManagerException` and `CrashedRoleException` are dealt with in the same way as other exceptions by the managers, i.e. if there is a exception handling interaction to deal with them, then this interaction is activated, otherwise these exceptions are raised in the callers of the DMI.

---

[3] A common exception from which all raised exception are extended from. In the worst case scenario, the common exception is `Exception`

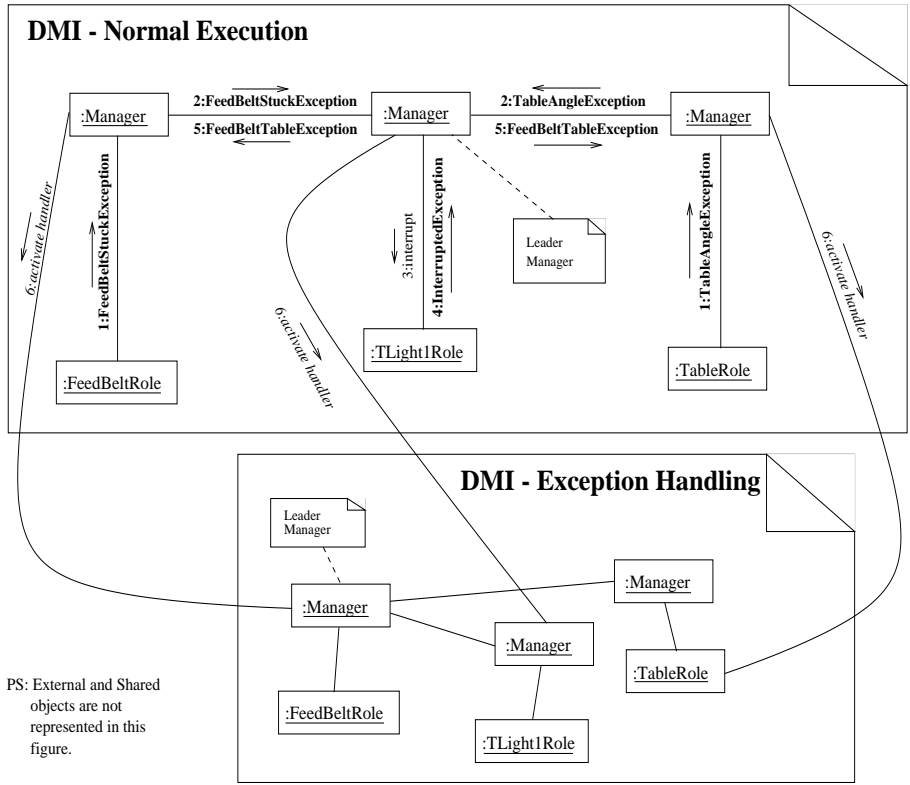

Figure 6: Dealing with Concurrent Exceptions

If the user of the framework wants to provide its own algorithm for deciding which exception is to be handled by all the roles, then the `Manager` class can be extended and a method called `exceptionHandling` must be provided. This method must return an exception that is derived from the `Exception` class. A list containing the exceptions that were raised by the roles is passed to the new exception handling method.

## 4.2 Discussion

The use of the framework presented in this section has helped us in disentangling interactions between objects from the definition of these objects. This facilitated the design and implementation of objects for the FZI Production Cell II case study, in the sense that objects that represent devices are only concerned with the basic operations of the devices. For example, in designing and implementing the robot object we had to consider only the operations that the robot can perform, e.g. operations to rotate the robot: *left*, *right*, and *stop*. Operations that are related to the environment of the robot are not designed/implemented in the robot object, e.g. the unloading of the table, or the loading of the press by the robot. Because these operations can vary depending on where the robot is installed, they are left to be implemented in a separate place, making the reuse of the robot object possible without modifications. DMIs suit this sort of strategy very well, with the additional benefit of enclosing and recovering possible failures that may happen during this kind of interaction.

The full Java implementation of the controlling software for the FZI Production Cell II is composed of 10 device/sensor objects, 23 exception objects, 61 role objects, and 7 other objects. Its design made use of the failure analyses and definitions from [Xu *et al.* 1999]. This controlling software was then applied to drive an FZI-provided graphical simulator to which we have added graphical means of representing how actions are progressing (either normally, or in response to exceptions that have been manually introduced via a fault-injector panel).

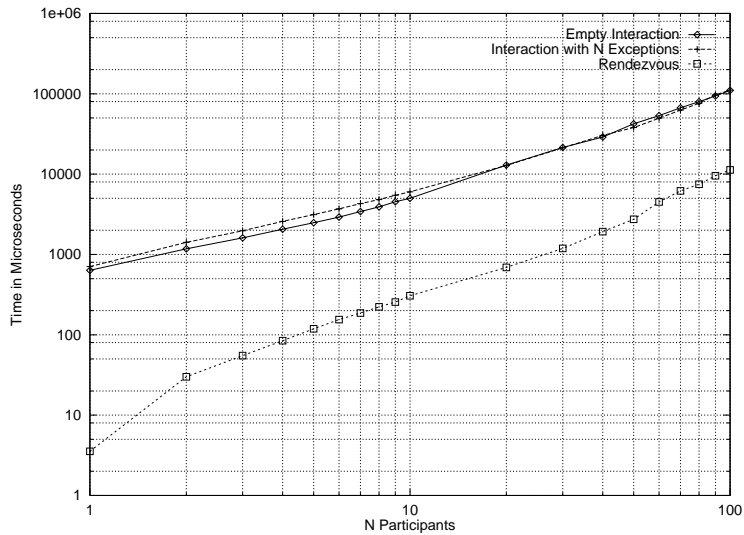

Figure 7: DMIs vs. Rendezvous

Figure 7 shows the performance costs of using our DMIs in the implementation of a system. The times were measured in a PC Pentium 200Mhz running Linux and Java version 1.1.6. All participants were running in the same Java Virtual Machine. We have measured the cost of an empty DMI and of a DMI in which exceptions are raised by all roles of that interaction. In the graph we compare these two executions of our framework with a simple rendezvous mechanism. Not surprisingly, the overhead of our mechanism is significant in comparison to that of a basic interaction mechanism, i.e. one that provides no support for the difficult problems of error containment and of handling even concurrent exceptions. Note also that increasing the number of participants does not cause the performance of our framework to degrade excessively, and it that scales in the same proportion as a simple rendezvous mechanism.

## 5 Conclusion

In this paper we have introduced the concept of dependable multiparty interactions (DMIs). We have also described how the DisCo language has been be extended in order to allow DMIs to be specified, and how to implement DMIs using an object-oriented API programmed in Java. The API that provides DMIs has been used to implement a controlling software for the FZI Production Cell II.

DMIs have proved in our experience to be a very effective means of describing cooperation between several participants even in the event of faults during the cooperation. We showed in this paper how we organize this kind of activities in an object-oriented fashion. Interactions between objects were extracted from the objects and enclosed by DMIs. This resulted in a very neat way of implementing basic objects to represent real devices in an industrial installation.

The way DMIs are activated by the underlying system is not taken into account in this paper. The properties used here suffice even if different scheduling mechanism are used or if the interactions are activated either synchronously or asynchronously.

**Acknowledgments**

We would like to thank our colleagues Robert Stroud, Jie Xu, Alexander Romanovsky and Ian Welch for their comments in the development of a framework for the CA action mechanism. We also thank several members of the ESPRIT Long Term Research Project 20072 on "Design for Validation" (DeVa). A. F. Zorzo is being supported by CNPq/Brazil - grant 200531/95.6.